\documentstyle[proceedings,epsf]{crckapb}

\begin{opening}
\title{THE COSMOLOGICAL PARAMETERS H$_0$ and $\Omega_0$}

\author{R. Brent TULLY}
\institute{University of Hawaii\\
           2680 Woodlawn Drive\\Honolulu, Hawaii 96822\\USA
\\ \\Proceedings IAU Symp. 183\\ Cosmological Parameters and the
Evolution of the Universe\\
(ed. K. Sato), Kyoto, Japan, 17-22 August, 1997.
}

\end{opening}

\begin{document}


{\it The first section discusses a recalibration of the luminosity--linewidth
technique and its use in a determination of ${\rm H}_0$.  The recalibration
introduces (i) new cluster calibration data, (ii) new corrections for
reddening as a function of inclination, and (iii) a new zero-point calibration
using 13 galaxies with distances determined via the cepheid period--luminosity
method.  It is found that H$_0=82\pm16$~km~s$^{-1}$~Mpc$^{-1}$ 
(95\% confidence).  

The second section concerns a dynamic measurement of $\Omega_0$.  The 
non-linear Least Action method can be used to reconstruct the orbits 
galaxies have followed to reach their current positions.  The three free
parameters of the models are M/L, age of the Universe, and a measure of
the possible extended nature of dark halos compared with the distribution
of light.  Models are constrained by 900 observed distances.  Best $\chi^2$
fits are for models with M/L values consistent with $\Omega_0=0.25\pm0.2$.  

The concept of `antibiasing' is introduced in an attempt to reconcile this
low estimate of $\Omega_0$ with the results from linear analyses.
It is argued that M/L values in elliptical-rich regions are a factor of $7$
larger than those in spiral-rich regions.  Hence, whereas it is commonly
thought that there is the bias that mass is more extended than light,
there may be a cosmologically important regime where there is the antibias
of mass more concentrated than light.  This possibility makes the
inferrance of $\Omega_o$ from $\beta$ unclear.}

\section{The Hubble Constant}

The Hubble Space Telescope (HST) has been used by three separate groups 
to detect cepheid variables in external galaxies, measure the luminosities
of those stars, and determine distances to the host nebulae
(Sandage et al. 1992; Freedman et al. 1994; Tanvir et al. 1995).  
The correlations between the
global luminosities of galaxies and rotation velocities (Tully \& Fisher
1977) are thought to produce good relative distances, but at the time of
my previous calibrations of the zero-points in various wavebands
(Pierce \& Tully 1988, 1992) there were only five nearby galaxies with 
distance estimates based on observations of cepheids with ground-based
telescopes.  Now there are 13 suitable galaxies
with cepheid distance determinations.  The HST cepheid programs are about
half completed.  An interum recalibration of the zero-point at $B,R,I$ 
bands is provided in this paper.

Two other improvements are made.  More optical, near-infrared,
and radio HI data is available so the calibration samples are larger and
more complete.  Also, the spectral coverage provided by $B,R,I$,$K^{\prime}$
photometry leads to a better specification of the obscuration
properties of galaxies.

\subsection{The Cluster Samples}

Malmquist bias caused by a magnitude limit (Malmquist 1920) can be 
neutralized if one has a suitable calibration sample.  There are three
requirements: one, the sample should be {\it complete} to an absolute 
magnitude limit fainter than the targets to be subsequently
studied; two, the intrinsic properties of the calibrators should be 
{\it statistically similar} to the targets; three, the {\it quality} 
of the 
observations should be similar.  With sufficient effort the completeness 
requirement can be met, and with sufficient care the quality of the 
observed material should be comparable between data sets.  It is 
impossible to be certain that the intrinsic properties of the calibrators
are truly representative but there can be some confidence in this 
proposition if several alternative calibration samples, representative of 
different environments, have consistent correlation slope and scatter
characteristics.

The current calibration uses 89 galaxies in three rather different clusters.
The obvious advantage of clusters is the possibility to achieve completeness
within a volume to an absolute magnitude limit.  The disadvantage is that
cluster galaxies may be systematically different from those in the field.
Two of the `clusters' used in this study are poor specimens as clusters
and were chosen because the environments are arguably the same as the 
`field'.  The Ursa Major Cluster contains 62 galaxies with 
$M_B^{b,i} < -16.5^m$, {\it no} ellipticals, only 10 S0-S0/$a$ types, and the 
rest are spirals and irregulars with normal HI content.  The cluster
has no central core, the radial velocity dispersion is only $\sim 150$~km/s
so a crossing time is $\sim 0.5 {\rm H}_0^{-1}$, and it can be inferred
that the region is in an early stage of collapse (Tully et al. 1996).
Excluding galaxies with $i<45^{\circ}$, ${\rm T}<{\rm S}a$, and 
interacting/peculiar, leaves a sample of 37 objects.
The Pisces `cluster' has been used frequently for distance scale studies
(Aaronson et al. 1986; Han \& Mould 1992) but it, too, is composed mainly of
HI-rich galaxies scattered about several insubstantial knots of early 
systems and may more properly be viewed as a prominent section of an 
intracluster filament (Sakai, Giovanelli, \& Wegner 1994).
Excluding galaxies with $i<60^{\circ}$, ${\rm T}<{\rm S}a$, interacting,
or with insufficient $S/N$ in the HI signal leaves a sample of 25 objects
with $M_B^{b,i} < -19^m$.
The third sample comes from the Coma Cluster, certainly a very rich and
dense environment quite unlike the field.  A calibration based only on 
this cluster would be subject to justifiable criticism.  However, it is 
found that  the properties of the correlations are compatible between 
this extreme environment and the others which is an argument that the 
methodology works over a wide range of environmental conditions.
The same exclusions as with Pisces
leaves a sample of 27 objects
with $M_B^{b,i} < -19.3^m$.
In sum, 89 galaxies are used in the present analysis that satisfy magnitude
completeness criteria.  About 20 galaxies are missing from a truly 
complete sample to the present limits because the HI detections currently
have insufficient $S/N$.

\subsection{Revised Reddening Corrections}

Photometry in four bands from $B$ to $K^{\prime}$, available for the UMa 
and Pisces
samples, make it possible to refine the corrections that must be made for
obscuration as a function of inclination.  The most sensitive tests use
deviations from mean color-magnitude
correlations involving $K^{\prime}$.  More inclined galaxies tend to be redder
because of obscuration.  Deviations from mean luminosity--linewidth 
correlations (more inclined galaxies tend to be fainter) can be used for
independent but somewhat less sensitive tests.

Reddening may be different for different classes of galaxies.  
Giovanelli et al. (1995) have presented a strong case for a luminosity
dependence among spiral systems.  There is enough information now 
available to specify the gross characteristics of this dependency.  The 
results from the 4-band color-magnitude analysis confirms the Giovanelli
et al. claim.  A simple linear dependence of the amplitude of reddening 
with magnitude is advocated at this time.  The enhanced corrections 
toward the bright end of the luminosity--linewidth relations causes 
a steepening of the correlations, particularly at the shorter wavelength
bands.  As a consequence, there is a considerably weaker color dependence 
of the slopes than experienced previously.  The new reddening corrections
are discussed by Tully et al. (1998).

\subsection{Zero-Point Calibration}

The neutralization of Malmquist bias is achieved with regressions that
accept errors in the distance independent variable (ie, the linewidths)
in a sample that is complete to a limit in the distance dependent variable 
(ie, magnitudes).  At a given magnitude, galaxies on either side of the 
mean  linewidth are equally accessible.  It is found that the slopes of 
the regressions on linewidth are compatible between the three cluster 
samples.  Hence, it seems justified to slide the separate cluster 
correlations in magnitude to find the relative distance differences that 
result in a common correlation.    A further constraint
is provided by the requirement that the relative distances agree between 
the measurements at different passbands:  the agreement between the 
$B,R,I$ filters is $0.03^m$ rms.  We
definite the {\it slopes} of the correlations and measure
the {\it relative distances} of the three clusters.  Pisces is determined 
to be $2.55^m$ farther away than UMa and Coma is determined to be 
$3.33^m$ farther away than UMa.

In the regime with $M_B^{b,i}<-16.5^m$,
there are now 8 galaxies with HST cepheid measurements to add to 5 
galaxies with ground-based distance determinations from cepheid
observations.  These 13 galaxies lead to luminosity--linewidth
correlations that are consistent with the combined cluster slopes with
even less scatter.  There is no contradiction to the proposition that 
the calibrators are similar to the cluster galaxies in properties and 
provide a reasonable zero-point determination to the magnitude scale.

\subsection{A Determination of H$_0$}

The $R$ band luminosity--linewidth relation for
the three clusters and calibrator data sets is displayed 
in Figure~1.  Relations with similar scatter exist at $I$ and
$K^{\prime}$ and, with greater scatter, at $B$.  The calibrators have not 
been observed at $K^{\prime}$ so, as yet, only relative distances can be
obtained at that band.
The Ursa Major Cluster has a distance 
modulus of 31.33, or 18.5~Mpc.  From the relative distances 
of the clusters, Pisces is at 60~Mpc and Coma is at 86~Mpc.  
Han \& Mould (1992) report the mean redshifts of these clusters in the
microwave background frame to be 4771~km~s$^{-1}$ and 7186~km~s$^{-1}$, 
respectively.
The Pisces and Coma regions are 
sufficiently distant that their velocities may be a close approximation
to the Hubble expansion.  The distances to these clusters indicate
H$_0=80$ and $84$~km~s$^{-1}$~Mpc$^{-1}$, respectively.

\begin{figure}
   \hspace{2cm} \epsfxsize=0.66 \textwidth \epsfbox{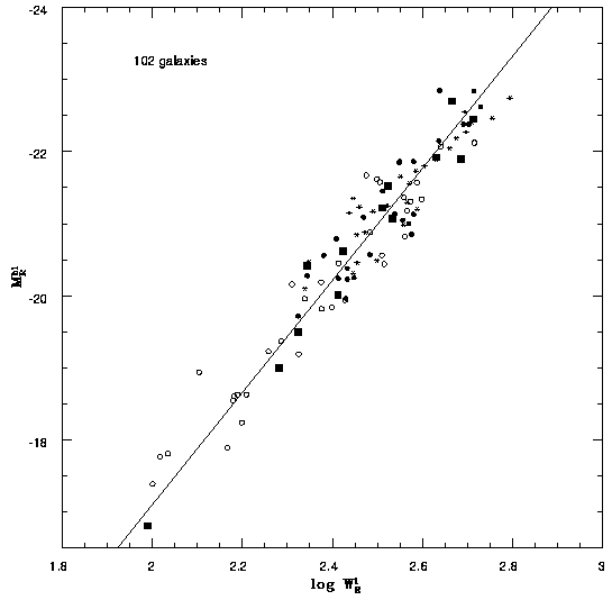}
\caption{$R$-band luminosity--linewidth relation.  The linewidth
parameter $W_R^i$ is a transformation from the observed HI 20\%
linewidth to a value statistically equal to twice the maximum rotation
velocity.  Symbols denote the various samples: {\it open circles}--UMa,
{\it filled circles}--Pisces, {\it stars}--Coma, {\it filled boxes}--
calibrators with cepheid distances.
}
\end{figure}

Statistical errors are estimated to be 8\% from the typical rms scatter
of $0.40^m$ and the numbers of objects in the 
samples.  There is about a 7\% uncertainty due to possible
peculiar velocities
of the two distant clusters.  The present 
zero-point assumes a modulus of 18.50 for
the Large Magellanic Cloud, uncertain to 10\%.
Systematic errors are
estimated to be $<10\%$ from the observation that the scatter 
in the luminosity--linewidth relations is $\sim0.4^m$ in case after case
with consistency in slopes between different environments.  Hence, 
systematics, apart from occasional blunders in photometry, are probably
not a large fraction of the scatter; ie, $< 1/2 \times 0.4^m \sim 0.2$.

These results are $\sim 13\%$ higher than other recent 
recalibrations of luminosity--linewidth correlations (Mould et al. 1996;
Giovanelli et al. 1997).  Lower values of H$_0$ are coming from the 
distances found by the HST measurements of cepheids which
tend to give host luminosities that are brighter 
at a given linewidth than found for the objects studied from the 
ground.  Maybe it is just small numbers.  There is about a
15\% increase of the distance scale measured in this study from the new
zero-point calibration.

On the other hand, there is a partially offsetting effect that comes about
through the new absorption corrections.  More luminous, highly inclined 
galaxies are made brighter.  The galaxies in Pisces and Coma tend to be 
brighter than in UMa and the samples are restricted to more inclined
systems in these clusters.  Hence, overall they receive larger corrections 
and the cummulative effect is to give them smaller distances by roughly
8\%.  

Compared with earlier measurements, the new cepheid calibration has 
caused an increase in distances, but the revised reddening corrections
has lead to smaller distances at larger redshifts.  The overall mix 
produces a lower H$_0$ than before but only by $\sim 7\%$.  It is 
sobering to see these systematics of order 10\% and to appreciate that 
even these matters we know about are not well resolved.  
A current  estimate of the Hubble
parameter is H$_0=82\pm16$~km~s$^{-1}$~Mpc$^{-1}$ (95\% confidence) 
from the
luminosity--linewidth method.  This error does not include uncertainty
in the zero-point attributable to the distance of the LMC.  There is 
work yet to be done.

\section{The Density Parameter}

The measurements of galaxy distances can give a map of {\it deviations}
from the Hubble expansion, something a few of us might find more interesting
than the Hubble Constant issue.  These deviant, or peculiar velocities 
are taken to arise from gravitational irregularities.  Flows of galaxies 
are responding to perturbations on scales of tens of megaparsecs.  Studies
of these flows provide the opportunity to make a kinematic 
measurement of the distribution of dark matter on megaparsec scales and
a determination of the cosmological density parameter 
$\Omega_0={\rm (mean~density)/(critical~density)}$.

\subsection{The Method of Least Action}

Shaya, Peebles, \& Tully (1995) describe a non-linear method for the
reconstruction of the orbits of a catalog of mass tracers.  The orbits 
follow trajectories that extremize the `action', the time integral of
the Lagrangian.  Boundary conditions on the orbits are the observed
current sky positions and redshifts and the initial conditions
that peculiar velocities were negligible.  A 
choice of mass-to-light ($M/L$) ratio transforms the observed luminosities of
the tracers into masses.  These masses tug on each other over the age of
the Universe.  
The Least Action model predicts a distance for each tracer.  These model
predictions are compared with observed distances in roughly 30\% of 
cases.  The match between model and observed distances provides a measure
of the quality of the model.

At present, three parameters define a model.  In the simplest case, there 
need only be two parameters.  If mass is very closely coupled to the light 
then it is sufficient to identify only an $M/L$ ratio and an age of the 
Universe to specify the time that the masses pull on each other.  A third
parameter can be introduced to address the complexity of biasing.  If
matter is more widely spread than the light then close-passing halos might 
begin to merge.  The effective masses will be reduced from the true masses.
A gimic to recover the true masses is to introduce a `softening parameter'.
Inside the scale of this parameter forces are reduced from the $r^{-2}$
expectation, so mass requirements are driven up.

\subsection{Application to Simulations and the Real Universe}

The Least Action procedure has been appplied to a standard Cold Dark 
Matter N-body simulation (Bond, Kofman, \& Pogosyan 1996) to test if 
the method recovers the known $\Omega_0=1$ property of the model.
If, naively, mass is assumed to be concentrated at the locations
of the halo tracers then the Least Action reconstruction fails to recover
all the mass.  The known mass is recovered,
however, if the softening parameter is taken to be $\sim 400$~km~s$^{-1}$.
Halos begin to inter-penetrate on roughly this scale.  

\begin{figure}
   \hspace{1cm} \epsfxsize=10.5cm \epsfbox{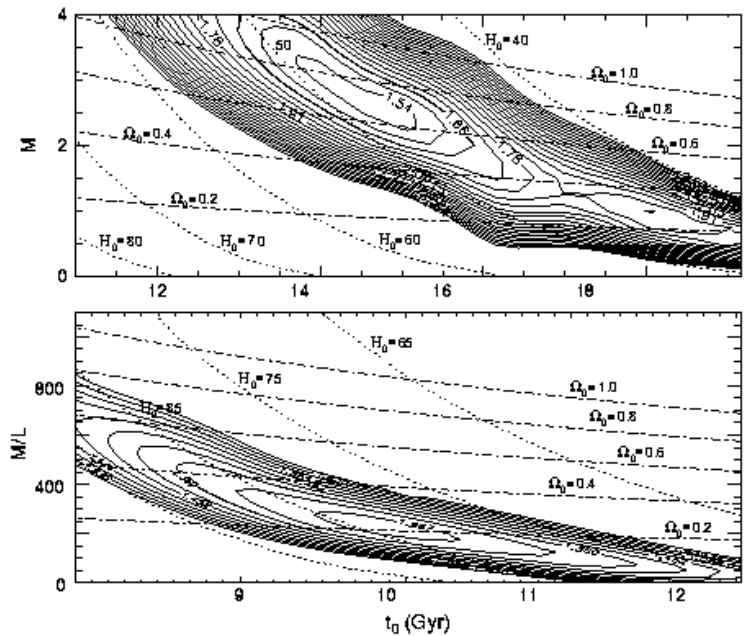}
\caption{{\it Top.} $\chi^2$ map of Least Action models of a
standard CDM N-body simulation.  The vertical mass scale is in arbitrary
units defined such that the true mass of the simulation is recovered 
with $M=4$.  The heavy contour is at the $2 \sigma$ level.  Curves of
$\Omega_0$ and H$_0$ are superposed.  In this case, the force softening
occurs on scales less than 250~km~s$^{-1}$ which is too small to optimally
recover the known parameters $M=4$, $\Omega_0=1$, H$_0=50$.
{\it Bottom.} Application to the real data.  Force softening occurs on
scales less than 300~km~s$^{-1}$ in this case, slightly less than the
case that produces the minimum $\chi^2$.
}
\end{figure}

The same procedure has now been applied to the real Universe.  A catalog
of 3030 galaxies within 3000~km s$^{-1}$ describes the distribution of 
matter.  Complex orbits are avoided by grouping galaxies in dense regions.
A total of 1323 groups/galaxies are identified that should never have 
intersected with each other.  The distribution of a combination of rich
clusters and sources selected from complete redshift surveys drawn from 
the InfraRed Astronomy Satellite provide a description of tidal
fields on scales greater than 3000~km s$^{-1}$.  The consistancy of the 
model distances are tested by comparison with 900 measured distances.
Most of these observed distances come from application of the 
luminosity--linewidth correlations discussed in the previous section
but 50 distances come from either cepheids, or surface brightness
fluctuations (Tonry, Ajhar, \& Luppino 1990), or planetary nebula 
luminosity functions (Jacoby, Ciardullo, \& Ford 1990).

In the simplest no-biasing case with mass distributed closely like the
light then the Least Action procedure results in best fits with very 
low densities: $\Omega_0=0.14\pm0.10$ (95\% confidence).  The error 
estimate is within the context of the no-bias assumption.  Systematic
uncertainties attributable to biasing are considerably larger. 
When the third parameter was admitted, $\chi^2$ fits improved
with increasing values of the softening parameter until an optimum
value of $\sim 400$~km s$^{-1}$ is reached.   In this case, 
$\Omega_0=0.25\pm0.2$.  
The uncertainty is larger, only approximate, 
and dominated by the vagaries of biasing.  The $\chi^2$ fits for
models near, but not quite, optimum are illustrated in Figure~2.
Though a lot more intercomparisons between N-body simulations and the
real data are required, it is reasonably convincing that the real world
is {\it not} described by the standard CDM model.

\subsection{Antibiasing}

The concept of biasing in the distribution of galaxies (Kaiser 1984)
arises from the reasonable possibility that starlight and dark matter
may not always be strongly correlated in position.  Dynamical 
interpretations of galaxy flows from linear analyses (Dekel, Bertschinger
\& Faber 1990; Hudson 1994; Davis, Nusser, \& Willick 1996)
measure the parameter $\beta$, with the relation to the physical parameter
of interest $\Omega_0=(\beta b)^{1.7}$ where $b$ is the
so-called `bias' parameter: 
$b={\rm (observed~fluctuations)/(true~mass~fluctuations)}$.
Experience with cosmological simulations and common 
sense give the appreciation that dark matter could be more widely 
distributed than the light of stars.  The great voids in the distribution
of galaxies may not be entirely empty of matter.  In this scenario,
the observed fluctuations in the galaxy distribution are a larger
fraction of the mean than the fluctuations in the total mass.  Hence,
$b>1$ and $\Omega_0 > \beta^{1.7}$.

At issue is why, by-and-large, the linear dynamical studies are claimed 
to give evidence for high values of $\Omega_0$, in the range $0.5-1$, 
(Dekel et al. 1990; da Costa et al. 1997; though $\beta$ values have 
been tending downward, see Riess et al. 1997) whereas the non-linear 
Least Action analysis persists in giving
values well below closure.  The non-linear analysis gives a direct measure
of masses, hence of $\Omega_0$ with a suitable normalization, and the
possibility of biasing in the measurement of masses is addressed within
the modeling, for example, by the 'force softening' trick.  The
premise with the linear analyses has been that the bias factor has
the property $b\ge1$.  Here, the possibility is raised that $b<1$, that
is, that mass fluctuations could be more concentrated than the light
in some circumstances.

The evidence comes from the Least Action modeling of regions of collapse,
particularly the infall region around the Virgo Cluster.  There are 
galaxies falling on first approach toward the cluster with line-of-sight
velocities with respect to the cluster of up to +900 and -600 km~s$^{-1}$.
These galaxies are not in the cluster; they are off the cluster in 
projection in the `southern extension'.  Distance measurements show 
conclusively that they are within the infall region around Virgo rather
than at the foreground or background `triple-value' location outside the
infall region.  The envelope of peculiar velocities as a function of 
angular separation puts a strong constraint on the mass of the Virgo 
Cluster of $\sim 1.3 \times 10^{15} M_{\odot}$ if the cluster is at a
distance of 16~Mpc.  This result implies 
$M/L \sim 1000 M_{\odot}/L_{\odot}$.  There is
evidence of similarly high $M/L$ requirements in other elliptical-rich
environments such as the Eridanus, Fornax, and Centaurus clusters.

There is the strong implication from the Least Action modeling that 
elliptical-rich environments have very large $M/L$ ratios.  On the other 
hand, the $M/L$ values for the vast bulk of the mass tracers cannot be 
nearly so high.  Indeed, if the elliptical-rich clusters are given the 
implicated high $M/L$ values then the residual mass requirements for 
the spiral-rich systems is modestly reduced.  Best $\chi^2$ fits are
coming in with $M/L \sim 150 M_{\odot}/L_{\odot}$.
Hence there is the extraordinary implication that 
$M/L({\rm E}) \sim 7 M/L({\rm S})$ where E means elliptical-rich 
environment and S means spiral-rich environment.  This result was
already found with the Virgo infall study by Tully \& Shaya (1984).
That earlier analysis used a non-linear but spherically symmetric
model.  It could have been feared that the approximation to
symmetry gave an insecure result but the non-parametric Least
Action model shows the earlier conclusion was sound.

Roughly 8\% of the $B$ light within 3000~km~s$^{-1}$ comes from 
elliptical-rich clusters.  However, with a factor of 7 boost,
some 40\% of the mass could be in these knots!  Roughly
10\% of galaxies are in rich clusters, but maybe a much larger
percentage of the collapsed matter is in these rich clusters.
If so, then the bias factor is $<1$ at least in high density
regions.  

It is not hard to think of reasons why $M/L$ values might be considerably
higher in dense clusters than in low density locations.  Stellar 
populations are older and redder.  Tidal stripping may have pulled
stars out of galaxies in clusters and there is evidence that there 
might be a comparable number of stars in the intra-cluster space as in
the galaxies.  There are more baryons in hot intracluster
gas than in stars in rich clusters.  Gas in the primordial cloud around
field galaxies will continue to rain down onto those galaxies perhaps 
even to the present epoch.  By contrast, if a galaxy falls into a cluster then 
the gas from its primordial cloud cannot reach that galaxy because of 
the Roche limit set by the tidal field of the cluster (Shaya \& Tully 1984).  
The gas supplies the intra-cluster medium not the galaxy.  

It is plausible that the bias factor is a complicated
function of environment, spatial scale, and passband.  Mass may be
more extended than light in low density regions and more concentrated than
light in high density regions.  Since the former effect has been called
`biasing', the latter effect can be called `antibiasing'.
The claim made here is that the latter
effect may be much more extreme than suspected, to the degree that it
puts the sense of the correction $\beta \rightarrow \Omega_0$ in doubt.

\section{References}


\noindent
Aaronson, M., Bothun, G., Mould, J.R, Huchra, J.P, Schommer, R.A., \&

Cornell, M.E.  1986, {\it Ap.J.}, {\bf 302}, 536.

\noindent
Bond, J.R., Kofman, L., \& Pogosyan, D.  1996, {\it Nature}, {\bf 380},
603.

\noindent
da Costa, L.N., Nusser, A., Freudling, W., Giovanelli, R., Haynes, M.P.,

Salzer, J.J., \& Wegner, G.  1997, astro-ph/9707299.

\noindent
Davis, M., Nusser, A., \& Willick, J.  1996, {\it Ap.J.}, {\bf 473}, 22.

\noindent
Dekel, A., Berschinger, E., \& Faber, S.M.  1990, {\it Ap.J.}, {\bf 364}, 349.

\noindent
Freedman, W.L. et al.  1994, {\it Nature}, {\bf 371}, 757.

\noindent 
Hudson, M.J.  1994, {\it M.N.R.A.S.}, {\bf 266}, 468.

\noindent
Giovanelli, R., Haynes, M.P., Salzer, J., Wegner, G., da Costa, L., \&
Freudling, 

W.  1995, {\it A.J.}, {\bf 110}, 1059.

\noindent
Giovanelli, R., Haynes, M.P., da Costa, L., Freudling, W., Salzer, J.J., \&

Wegner, G. 1997, {\it Ap.J.}, {\bf 477}, L1.

\noindent
Han, M.S. \& Mould, J.R. 1992, {\it Ap.J.}, {\bf 396}, 453.

\noindent
Jacoby, G.H, Ciardullo, R., \& Ford, H.C.  1990, {\it Ap.J.}, {\bf 356}, 332.

\noindent 
Kaiser, N.  1984, {\it Ap.J.}, {\bf 284}, L9.

\noindent
Malmquist, K.G.  1920, {\it Medd. Lund Astron. Obs.}, Ser. 2, No. 22.

\noindent
Mould, J., Sakai, S., Hughes, S., \& Han, M.  1996, in {\it The 
Extragalactic

Distance Scale}, A.S.P. Series, ed. M. Livio.

\noindent
Pierce, M.J., \& Tully, R.B.  1988, {\it Ap.J.}, {\bf 330}, 579.

\noindent
Pierce, M.J., \& Tully, R.B.  1992, {\it Ap.J.}, {\bf 387}, 47.

\noindent
Sakai, S., Giovanelli, R., \& Wegner, G.  1994, {\it Ap.J.}, {\bf 108}, 33.

\noindent
Sandage, A., Saha, A., Tammann, G.A., Panagia, N., \& Macchetto, F.D.

1992, {\it Ap.J.}, {\bf 401}, L7.

\noindent
Shaya, E.J., \& Tully, R.B.  1984, {\it Ap.J.}, {\bf 281}, 56.

\noindent
Shaya, E.J., Peebles, P.J.E., \& Tully, R.B.  1995, {\it Ap.J.}, {\bf 454}, 15.

\noindent 
Tanvir, N.R., Shanks, T., Ferguson, H.C., \& Robinson, D.T.R.  1995, 

{\it Nature}, {\bf 377}, 27.

\noindent
Tonry, J.L., Ajhar, E.A., \& Luppino, G.A.  1990, {\it A.J.}, {\bf 100}, 1416.

\noindent
Tully, R.B., \& Fisher, J.R.  1977, {\it Astron. Astrophy.}, {\bf 54}, 661.

\noindent
Tully, R.B., Pierce, M.J., Huang, J.S., Saunders, W.,  Verheijen, M.A.W., 

Witchalls, P.L.  1998, to be submitted to {\it A.J.}.

\noindent
Tully, R.B., \& Shaya, E.J.  1984, {\it Ap.J.}, {\bf 281}, 31.

\noindent
Tully, R.B., Verheijen, M.A.W., Pierce, M.J., Huang, J.S., \&
Wainscoat, 

R.J.  1996, {\it A.J.}, {\bf 112}, 2471.

\end{document}